\documentclass[twocolumn,prb,showpacs,floatfix,aps]{revtex4} %
\usepackage{amsmath,eucal}
\usepackage{graphicx}
\usepackage{pifont}
\usepackage{verbatim}
\usepackage[latin1]{inputenc}
\usepackage{calc}
\usepackage{subfigure}
\usepackage{rotating}

\begin{document}

\title{Cumulene Molecular Wire Conductance from First Principles}

\author{J. Prasongkit$^{1}$,   A. Grigoriev$^{1}$, G. Wendin$^{2}$ and Rajeev Ahuja$^{1,3}$}
\affiliation{$^{1}$ Division of Materials Theory, Department of Physics, Uppsala University, 75121 Uppsala, Sweden \\
$^{3}$ Department of Microtechnology and Nanoscience-MC2, Chalmers University of Technology, S-41296 G$\ddot{o}$teborg, Sweden\\
$^{2}$ Applied Material physics, Department of Materials and Engineering, Royal Institute of Technology (KTH), 10044, Stockholm, Sweden}

\date{\today}

\begin{abstract}

We present first principles calculations of current-voltage characteristics (IVC) and conductance of Au$(111)$:S$_2$-cumulene-S$_2$:Au$(111)$ molecular wire junctions with realistic contacts. The transport properties are calculated using full self-consistent $ab$ $initio$ nonequilibrium Green's function density-functional theory methods under external bias. The conductance of the cumulene wires shows oscillatory behavior depending on the number of carbon atoms (double bonds). Among all conjugated oligomers, we find that cumulene wires with odd number of carbon atoms yield the highest conductance with metallic-like ballistic transport behavior. The reason is the high density of states in broad lowest unoccupied molecular orbital levels spanning the Fermi level of the electrodes. The transmission spectrum and the conductance depend only weakly on applied bias, and the IVC is nearly linear over a bias region of ${\pm}$1V. Cumulene wires are therefore potential candidates for metallic connections in nanoelectronic applications.

\end{abstract}


\maketitle

\section{Introduction}

Molecular nanowires and atomic chains are actively explored as novel conductors for nanoelectronic devices. Early on, cumulenes carbon chains were proposed as ideal molecular wires \cite{Beck93}, and the interest in the structure persists \cite{Milani,YangPCCP}. Subsequent work investigated the transport properties of cumulene wires directly connected to jellium  \cite{Lang98,Langtr} or atomically structured \cite{Palacios:2002kq,Transi} electrodes via carbon double bonds. For both types of electrodes, the conductance was found to vary in an oscillatory manner with the number of carbon atoms in the finite chain. Nevertheless, the transmission not only depends on the properties of the wire, but also on the contacts to the electrodes. For example it was understood \cite{Langtr, Palacios:2002kq} that the amount of charge transferred from the electrodes to the molecular wire chemically bonded to the metal electrodes, determines the character of the conductance and its length dependence. The charge transfer from the electrodes to the cumulenes provides doping of wires\cite{Langtr}. Since reliable methods of purification of cumulenes have been established\cite{Bildstein04,Suzuki07}, we reconsider the transport properties of these molecules using a realistic metal-molecule interface model. 

In this paper we report full self-consistent ab initio calculation of electron transport through the cumulene molecular wires =$($C=$)_{n}$ with $n$=4-9 connected to Au$(111)$ surfaces via thiolate bonds. Unlike the previous studies of the cumulenes \cite{Lang98, Langtr, Palacios:2002kq}, where the cumulative double bond order was supported in the metall\emph{A}cumulene form, with the metal atom at the electrode surface incorporated into the carbon double-bond framework, we here consider the full atomic structure of \textquoteleft{}simple\textquoteright{} organic cumulenes of type R$_{\text{2}}$C=(C=)$_{\text{n-2}}$CR$_{\text{2}}$, forming metall\emph{O}cumulene complex with the metal surface. Metallic-like conductance of another form of carbon chains with alternating single and triple bonds, polyynes, was recently studied by Crljen and Baranovi$\acute{\mbox{c}}$ \cite{poly} using thiloate contacts; however, such a realistic model for the metall\emph{O}cumulene wires was never investigated, because the molecule itself was considered to be difficult to isolate in macroscopic quantity.

\section{Computational method}

The calculations are performed using density functional theory-based nonequilibrium Green's function (NEGF) transport theory as implemented in the TranSIESTA simulation package \cite{Transi}. We used local-density approximation (LDA) for the exchange correlation potential. For carbynes, LDA performed better than generalized gradient approximation (GGA) for calculating the structures \cite{a}. Using LDA for relaxation and reconstruction on Au(111) surface has been extensively discussed in previous literature \cite{pi}. 
For the transport properties one can expect the LDA to perform quite well due to the metallic character of the junction \cite{poly}.
Core electrons were modeled with Troullier-Martins \cite{TM} soft norm-conserving pseudopotentials and the valence electrons were expanded in a basis set of local orbitals \cite{Grigoriev06}. The transmission spectrum, which gives the probability for electrons with incident energy $E$ to be transferred from the left to the right electrode, is calculated from \cite{formu}
{\small{
{\begin{equation}
T(E,V)=\mathbf{tr}\left[\Gamma_{R}(E,V)G_{C}(E,V)\Gamma_{L}(E,V)G_{C}^{+}(E,V)\right]
\end{equation}}
}}where $G_{C}$ is the Green's function of the central region and $\Gamma_{L/R}$ is the coupling matrix of Left/Right electrode, respectively. The integration of the transmission spectrum yields the electric current 
{\small{
{\begin{equation}
 I(V)=\int_{\mu_{L}}^{\mu_{R}}{T(E,V)\left\{f(E-\mu_{L})-f(E-\mu_{R})\right\}\,dE}
\end{equation}}
}}
where $\mu_{L}=-V/2\;(\mu_{R}=V/2)$ is the chemical potential of the left(right) electrode.

\begin{figure}[!ht!]
\begin{minipage}[t]{1.0\linewidth}
\begin{center}
\includegraphics[width=8.15cm, height=5.5cm]{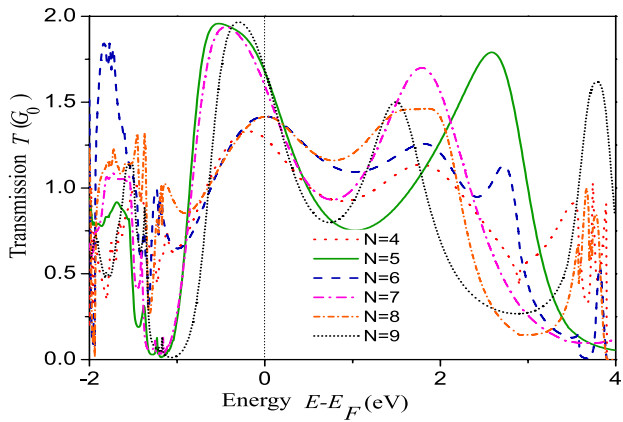}
\end{center}
\end{minipage}
\caption{The zero bias electron transmission spectra of cumulene systems consisting of $n$=4-9 carbon atoms.}
\label{fig:transmissions}
\end{figure}

In our calculations, free thiol-capped cumulenes were first optimized using the SIESTA package \cite{siesta}. The bond lengths and geometrical structures obtained in all the performed calculations are similar to those of unsubstituted cumulenes \cite{lenght}. The minimum-energy conformation of even-number wires, is planar, whereas the terminal groups of odd-number wires are mutually perpendicular. A system of a cumulene wire coupled to Au(111)-$(3\times3)$ surface electrodes via thiolate bonds was fully optimized \cite{Grigoriev06}. The adsorption geometry has been determined by positioning thiolate capped cumulenes above four adsorption sites, on-top, bridge, fcc and hcp on Au surfaces at a favorable Au-S bonding distance \cite{distance}. During relaxation sulfur atoms move to the bridge sites. Then the metal-molecule-metal sandwich was constructed, and fully optimized. The Au-S equilibrium distance is 2.20 \AA. 

To establish a firm ground concerning cumulene adsorption geometry, cumulene molecules on Au$(111)$ for all $n$ considered were optimized by using the VASP \cite{vasp} simulation package with LDA potential, following the computational procedure described in Ref. \onlinecite{Mankefors03}; then the metal-molecule-metal sandwiches were constructed. We obtained identical geometries, as compared to SIESTA calculations, with marginal differences noticed only for the longest molecules. It is instructive to compare the sulfur adsorption geometry obtained in this work to the one obtained in Ref. \onlinecite{Mankefors03}. The previously calculated tendency was for sulfur atoms to move from in-hollow to bridge and then to on-top position on Au$(111)$ when the sulfur density and/or hydrogenation density increases. This perfectly matches the present findings where two sulfur atoms brought together with bonds to a common carbon atom (but located relatively far from other Au:S assemblies) slide toward bridge positions from any initial adsorption geometry. For the sandwiched molecule, the bonding energy was found to be ~1.7 eV per S atom for even $n$ and ~1.86 eV per S atom for odd n. For $n$=8,9 these energies becomes 0.1 eV smaller.

\begin{figure}[!ht!]
\begin{minipage}[t]{1.0\linewidth}
\begin{center}
\includegraphics[width=8.15cm]{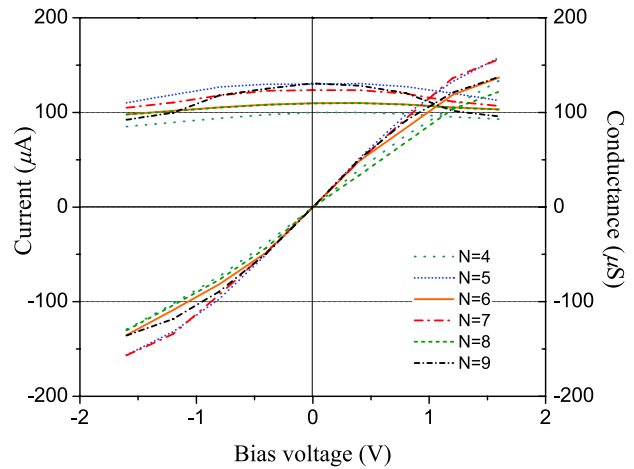}
\end{center}
\end{minipage}
\caption{The IVCs and corresponding differential conductance of the cumulene systems with 4-7 carbon atoms in the bias region from -2 to 2 V.}\label{fig:IV}
\end{figure}

\begin{figure}[!hb]
\begin{minipage}[t]{1.0\linewidth}
\begin{center}
\includegraphics[width=8.15cm]{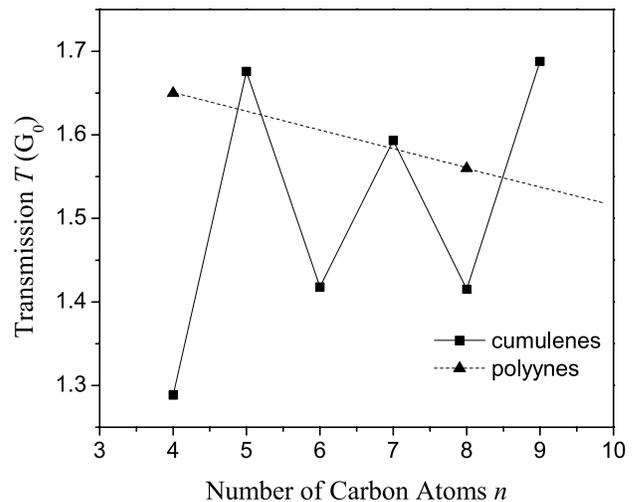}
\end{center}
\end{minipage}
\caption{Zero-bias transmission $T(E=0,n)$ through the carbon chains.} 
\label{fig:conductanceoscillations}
\end{figure}

\begin{figure*}
\begin{minipage}[t]{1.0\linewidth}
\begin{center}
\includegraphics[width=15.5cm]{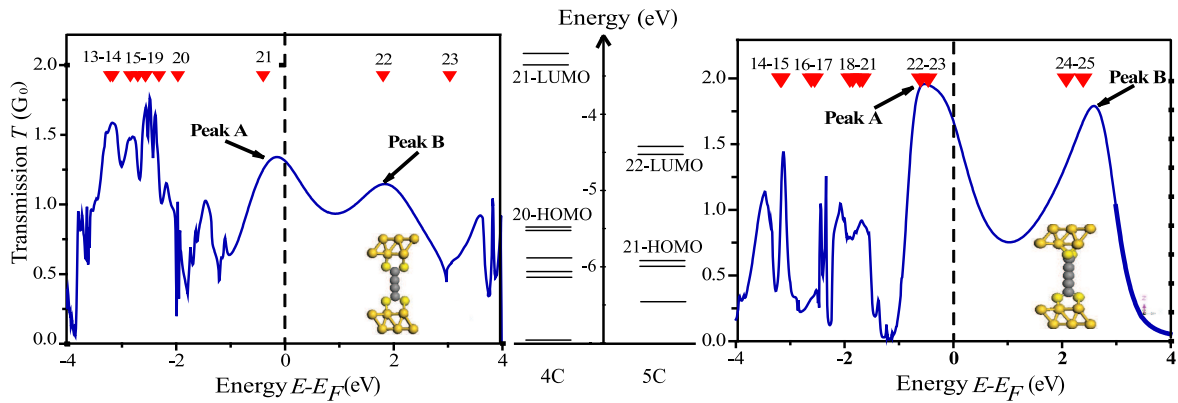}
\end{center}
\end{minipage}
\caption{Transmission spectra and molecule projected self-consistent Hamiltonian (MPSH) of cumulene systems (a) tetraene (b) pentane at zero bias, as compared to energy levels of corresponding radicals. Atoms in central region included in the self-consistent cycle are shown in the inset. The broad resonances are marked as peak A and B at the first peak below and above the Fermi level, respectively.
}
\label{fig:transmissionsdetails}
\end{figure*}

\section{Results and Discussion}

Fig.~\ref{fig:transmissions} shows the zero bias electron transmission spectra ($T$) for the studied metal-molecule-metal systems, while the corresponding IVCs and differential conductances are shown in Fig.~\ref{fig:IV} in the bias region from -2 to 2 V. The zero bias conductance in Fig.~\ref{fig:conductanceoscillations} clearly demonstrates the oscillating behavior of the conductance of odd/even-$n$ cumulene chains. 

To examine the results in detail, we concentrate on the tetraene and pentaene chains. Fig.~\ref{fig:transmissionsdetails} shows transmission spectra and molecule projected self-consistent Hamiltonian (MPSH) eigenvalues of the cumulene systems (a) tetraene (b) pentaene at zero bias. Atoms in the central region included in the self-consistent cycle are shown in the lower inset. The MPSH eigenvalues are in fair agreement with the energies at the transmission resonance peaks. The main features of transmission spectra are the broad resonances below and above the Fermi level, marked as ``peak A" and ``peak B", respectively. Comparing the MPSH states with molecular orbitals (we use deprotected radical for clarity) we conclude that the density of states contributing to the peaks A and B arises from the unoccupied levels, and that the LUMO level dominates transmission channels. Undoubtedly, the high value of the conductance of cumulenes is to be  expected because of the delocalized nature of the LUMO.

Oscillatory behavior of conductance with the number of carbon atoms in the chain is evident from Fig.~\ref{fig:conductanceoscillations}, with odd-$n$ carbon atom wires giving a higher conductance than even-$n$ ones. All cumulenes demonstrate high conductance, higher than other conjugated polymers, and for the odd-$n$ cumulene wires the conductance is even higher than polyynes \cite{poly} (see Fig.~\ref{fig:conductanceoscillations}). The high value of the conductance results from the high density of states at the Fermi level, with the position of the LUMO aligned with Fermi level, as clearly seen in Fig.~\ref{fig:transmissionsdetails}. While the conductance of polyynes slowly decreases with increasing wire length, the conductance of odd-$n$ cumulenes does not show any pronounced dependence on the molecular length, which is a clear signature of ballistic conductance. Because of ballistic transport, backscattering in molecular wires should be negligible, with little energy dissipation, with finite resistance attributed to the effect of backscattering from the contact region (constriction and interface structure). Previous work \cite{Langtr} on cumulene systems with jellium electrodes has found Schottky-like barriers at the metal-wire contacts. Thus, we conclude that the reason for deviation from the maximum conductance predicted by Landauer theory is backscattering from the contact between the electrode and the molecule, emphasizing the importance of \emph{ab initio} studies of interface chemistry and electron structure.

\begin{figure}[!b]
\begin{center}
\includegraphics[width=8.15cm]{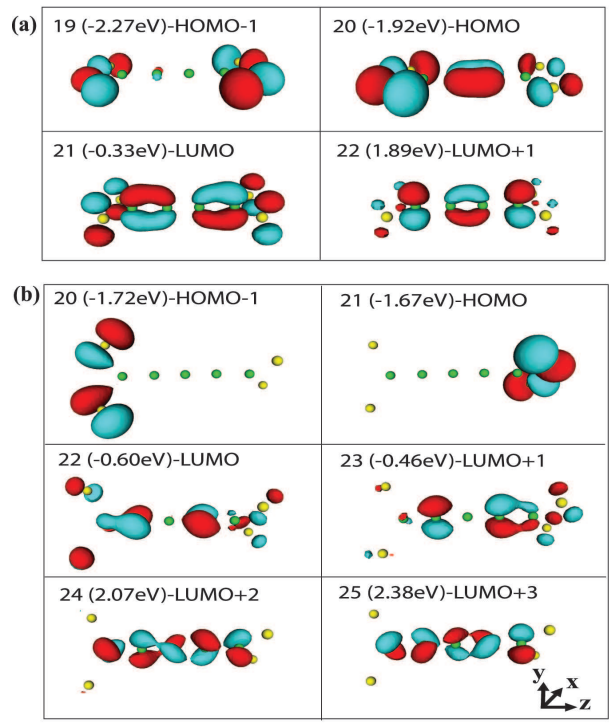}
\end{center}
\caption{MPSH states of (a) tetraene and (b) pentaene systems around the Fermi level; the corresponding radical orbital is referenced according to the level structure on Fig.~\ref{fig:transmissionsdetails}.}
\label{fig:orbitals}
\end{figure}

In the jellium electrode model, oscillatory behavior can be explained \cite{Langtr} by the different occupation of molecular orbitals. In that model the  highest occupied molecular orbital (HOMO) in the even-$n$ free chain is half occupied, which leads to misalignment of the molecular HOMO with the Fermi level of the electrodes, whereas the HOMO in odd-$n$ free chains is fully occupied, which leads to partial discharge of the molecular HOMO and alignment with electrodes Fermi level. Thus, even(odd)-$n$ chains have low(high) density of states at the Fermi level. Comparing with our results with thiolate-capped cumulenes chemisorbed onto gold electrodes, we find that the molecular level alignment has changed because of strong hybridization between molecular and electrode surface states coupled by thiolate bonds. The position of the LUMO level lies below the Fermi level at zero bias for both even and odd numbered cumulene wires. Charge transfer from the electrodes to the molecule can further partially fill the LUMO. Therefore, the conductance of even-$n$ wires is still quite high at zero bias. Comparing transmission spectra for odd-$n$ and even-$n$ carbon number wires, we see that the main peaks near the Fermi level are rather similar, but the peak magnitudes and widths are different. 

\begin{figure}[!b]
\begin{minipage}[t]{1.0\linewidth}
\begin{center}
\includegraphics[width=8.15cm]{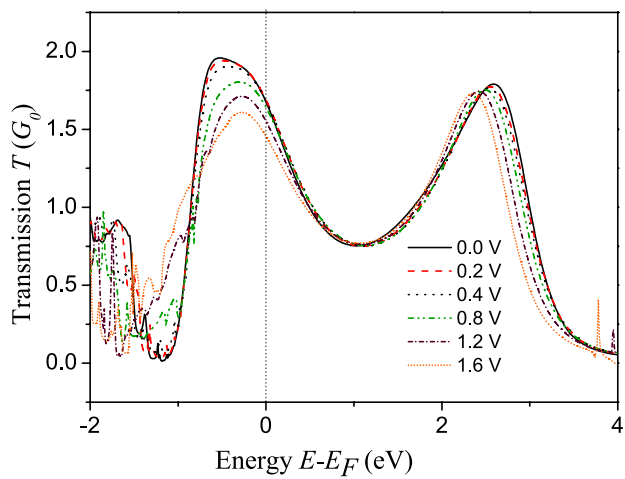}
\end{center}
\end{minipage}
\caption{Transmission spectra of pentaene system as a function of bias voltage. Energies obtained from the average electrochemical potential of the electrodes.}
\label{fig:T(V)}
\end{figure}

To elucidate the origin of this behavior, we investigate how the individual states of cumulene wires contribute to the conductance. MPSH states on the cumulenes are therefore analyzed to relate transmission spectra to molecular orbitals. As indicated by the MPSH eigenvalues in Fig.~\ref{fig:transmissionsdetails}, the two main transmission peaks (marked `A' and `B') of tetraene are mainly contributed by states 21 and 22, while the peaks of pentaene take their origin from states 22, 23, 24 and 25. The corresponding MPSH states, which are mainly responsible for the transmission around the Fermi level, are shown in Fig.~\ref{fig:orbitals}. It is found that common features of these orbitals are their significant delocalization along the backbone of the wire. Although the shape of peak A and peak B of even-$n$ and odd-$n$ wires are rather similar, the transmission channels inside the molecular wires are different. We find that $\pi_{x}$ states of the LUMO level located at peak A and peak B of tetraene have no pronounced appearance (Fig.~\ref{fig:orbitals}a, state 21 and 22), while those of pentaene show both $\pi_{x}$ and $\pi_{y}$ states (Fig.~\ref{fig:orbitals}b, states 22, 23, 24 and 25). Thus the occupation of the tetraene LUMO+1 (state 22) upon adsorption of the molecule between metal electrodes becomes unfavorable, due to the close symmetry of this state and lower-lying state 21, which keeps the state far above Fermi level. The $\pi_{y}$ state in tetraene appears at MPSH state 20, which is far away from the Fermi level. This makes the number of transmission eigenchannels around the Fermi level of odd-$n$ wires higher than in even-$n$ wires. This physical picture of electrode-induced interaction and hybridization of states upon chemisorption clarifies the origin of the oscillatory conductance in cumulenes wires.

Due to its symmetric structure, the \emph{I}-\emph{V} curve and conductance of cumulenes are symmetric with bias polarity. As the applied voltage is increased, the conductance of cumulenes slowly decreases. This result can be explained by plotting transmission spectra as a function of bias voltage, as shown in Fig.~\ref{fig:T(V)}. As positive bias is applied, the shapes of transmission spectra remain similar, whereas amplitudes of peak A decrease simultaneous with the decrease gap between peak A and peak B. The change of its position and magnitude relative to the electrochemical potentials of each electrode results in a gradual drop of conductance. The obtained curves demonstrate metallic IVCs with linear increase in bias range from -0.9 to 0.9 V of odd-number wire, and from -0.7 to 0.7 V of even-number wire. Above those values of bias voltage, \emph{I}-\emph{V} curves start to deviate from linearity. In the linear regime, the slopes of the \emph{I}-\emph{V} curves give the conductance (\emph{G}) of the cumulene systems at zero bias, which is 1.29 for $n=4$, 1.68 for $n=5$, 1.42 for $n=6$, 1.59 for $n=7$, 1.42 for $n=8$, 1.69 for $n=9$ in units of quantum conductance ${(G_{0})}$. Since our electrodes are very different from jellium electrodes, the obtained magnitude of conductance is slightly lower than obtained in Ref.~\onlinecite{Lang98}. Additionally, the conductance change upon stretching of the wire was found to be of the order of 0.1 ${(G_{0})}$ for 1 $\AA \; $change in the inter-electrode distance.

\section{Summary}

We have presented calculated IVCs of the cumulene wire junctions demonstrating ballistic-like transport. The high metallic-like conductance varies in an oscillatory manner with odd-even $n$: odd-$n$ ${\approx}$1.7 ${(G_{0})}$, even-$n$ ${\approx}$1.4 ${(G_{0})}$. This study indicates that cumulenes are worthy of further theoretical and experimental research, with potential to become useful one-dimensional molecular metallic wires for nano-scale electronic devices. We emphasize the decisive role of the anchoring group for adsorbed molecular structure and properties.

\section{Acknowledgements}
J.P. acknowledges support from the Royal Thai Government. A.G. and R.A. gratefully acknowledge financial support from Carl Tryggers Stiftelse f\"{o}r Vetenskaplig Forskning and U3MEC, Uppsala. G.W. acknowledges support from the EU ICT-NANO projects NABAB and SINGLE. The calculations were performed at the high performance computing centers UPPMAX within the Swedish National Infrastructure for Computing.

\end{document}